# On Generating Cancelable Biometric Template using Reverse of Boolean XOR

*Manisha, Nitin Kumar*
*Department of Computer Science and Engineering,*
National Institute of Technology, Uttarakhand,
Srinagar Garhwal, Uttarakhand, India
manisharawatphd@nituk.ac.in, nitin@nituk.ac.in

*Abstract*—Cancelable Biometric is repetitive distortion embedded in original Biometric image for keeping it secure from unauthorized access. In this paper, we have generated Cancelable Biometric templates with Reverse Boolean XOR technique. Three different methods have been proposed for generation of Cancelable Biometric templates based on *Visual Secret Sharing scheme*. In each method, one *Secret image* and *n*-1 *Cover images* are used as: (*M1*) One original Biometric image (Secret) with n- 1 randomly chosen Gray Cover images (*M2*) One original Secret image with *n*-1 Cover images, which are Randomly Permuted version of the original Secret image (*M3*) One Secret image with *n*-1 Cover images, both Secret image and Cover images are Randomly Permuted version of original Biometric image. Experiment works have performed on publicly available ORL Face database and IIT Delhi Iris database. The performance of the proposed methods is compared in terms of Co-relation Coefficient (Cr), Mean Square Error (MSE), Mean Absolute Error (MAE), Structural Similarity (SSIM), Peak Signal to Noise Ratio (PSNR), Number of Pixel Change Rate (NPCR), and Unified Average Changing Intensity (UACI). It is found that among the three proposed method, *M3* generates good quality Cancelable templates and gives best performance in terms of quality. *M3* is also better in quantitative terms on ORL dataset while *M2* and *M3* are comparable on IIT Delhi Iris dataset.

## I. INTRODUCTION

In Cancelable Biometric template generation process, original Biometric is distorted by applying one or more methods to such an extent that the original Biometric becomes extremely difficult to recover [7], [8]. A simple overview of this template generation process is demonstrated in Fig 1. Cancelable Biometric template generally possess four characteristics i.e. (i) Different users must be allocated with different Cancelable Biometric template (ii) Cancelable Biometric template should not reveal any information about individual's original Biometric (iii) Applying reverse engineering process on Cancelable Biometric templates, intruder should not succeed in retrieval of original Biometric (iv) Performance of the Cancelable Biometric Based applications should be same as original Biometric based applications. In literature, several methods have been suggested by various researchers for generation of such templates. Recently, a comprehensive survey of these methods is carried by Manisha and Kumar [1]. In this research work, Cancelable Biometric template generation methods are broadly categorized in to six categories based on following methods: Cryptography, Transformation, Filter, Hybrid, Multimodal and Other methods. For better understanding of various methods under these categories, please refer [1], [3] and [6]. Recently, Secret sharing methods [5] have gained wide popularity for data hiding. One such type of Secret sharing scheme is (*k*, *n*)-Visual Cryptography [4] in which at least *k* out of *n* images are required to successfully recover the original image. On the other hand in *(n, n)*-Visual Cryptography in which exactly *n* shares are required to recover the original image. In this work, we have exploited Visual Secret Sharing scheme for Cancelable Biometric template generation.

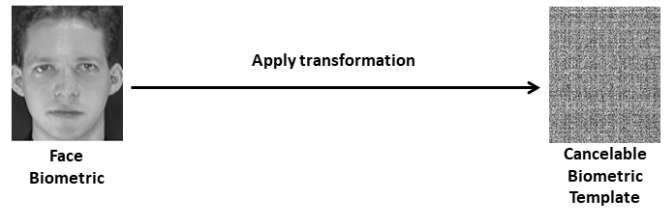

Fig. 1. Cancelable Biometric template generated from original Biometric.

## II. CONTRIBUTION

Here, we propose a novel scheme for Cancelable Biometric template generation using Visual Secret Sharing scheme in which Boolean XOR operation is performed as an intermediate step. The major contribution of our proposed work is three fold: (i) Three different methods *M1*, *M2* and *M3* have been proposed for generating Cancelable Biometric template (ii) Visual Secret Sharing scheme is used at storage and authentication phase in a distributed manner and (iii) Better security of Cancelable templates in terms of various performance measures.

## III. PROPOSED WORK

To generate Cancelable Biometric template(s) for a Biometric image, we need one or more other images which shall be combined with the original Biometric image. The original image is called *Secret image* and the other images are called

*Cover images*. Here, the Cover images are chosen in different ways based on which we have proposed three different methods i.e. (*M1*) One original Biometric image (S*ecret*) with n-1 Randomly chosen gray images as Cover images (*M2*) One original image (*Secret*) with *n*-1 randomly permuted version of original image as Cover images (*M3*) All images (*Secret and Cover images*) are randomly permuted version of original Biometric image.

TABLE I.
**Acronyms Used**

| S | Secret image | C | Cover images |
|---|---|---|---|
| SS | Secret Shares | NS | Noisy Shares |
| Cr | Correlation | TS | Temporary Shares |
| MSE | Mean Square Error | LBR | Left Bitwise Reversal |
| MAE | Mean Absolute Error | RBR | Right Bitwise Reversal |
| ATS | Authentication phase Temporary Shares | ANS | Authentication phase Noisy Shares |
| PSNR | Peak Signal to noise ratio | NPCR | Number of Pixel Change Rate |
| SSIM | Structral Similarity | UACI | Unique Average Changing Intensity |

Motivated by the research work of Deskmukh et al. [2], here we propose the step by step procedure for generation of Cancelable Biometric templates as shown in Tables II and III respectively for Enrollment and Authentication phase. These two algorithms are same for *M1*, *M2* and *M3* except the fact that the set of original Secret image and Cover images are supplied in a different manner based on a *M1*, *M2*, *M3* methods. The detailed working of these three methods is explained in Subsections III (A), III (B) & III (C) respectively.

TABLE II.
**Algorithm for Enrollment**

**Input**: one Secret image $S_1$ and *n*-1 Cover images $C_1, C_2, C_3,...,C_{n-1}$
**Output**: *n* secret shares $SS_1, SS_2, SS_3,...,SS_n$
Steps:
1. Generate Temporary Shares $TS_1, TS_2, TS_3,...,TS_n$
(i) $TS_1 = S_1$
(ii) $TS_i = C_{i-1} \oplus TS_{i-1}$   for $i = 2$ to $n$
2. Generate Noisy Shares $NS_1, NS_2, NS_3,..., NS_n$
(i) $NS_1 = TS_n$
(ii) $NS_i = TS_i \oplus NS_{i-1}$   for $i = 2$ to *n*-1
(iii) $NS_n = TS_1 \oplus NS_{n-1}$
3. Generate Secret Shares $SS_1, SS_2, SS_3,..., SS_n$
   $SS_i = LBR(NS_i)$   for $i = 1$ to $n$

TABLE III.
**Algorithm for Authentication**

**Input**: *n* Secret Shares $SS_1, SS_2, SS_3, SS_4,..., SS_n$
**Output**: one Secret image $S_1$ and *n*-1 Cover images $C_1, C_2, C_3,...,C_{n-1}$
Steps:
1. Generate Noisy Shares $NS_1, NS_2, NS_3,...,NS_n$
   $NS_i = RBR(SS_i)$   for $i = 1$ to $n$
2. Generate Temporary Shares $TS_1, TS_2, TS_3,...,TS_n$
(i) $TS_1 = NS_n \oplus NS_{n-1}$
(ii) $TS_i = NS_i \oplus NS_{i-1}$   for $i = 2$ to *n*-1
(iii) $TS_n = NS_1$
3. Generate Secret image $S_1$ and Cover images $C_1, C_2, C_3,...,C_{n-1}$
(i) $S_1 = TS_1$
(ii) $C_{i-1} = TS_i \oplus TS_{i-1}$   for $i = 2$ to *n*-1

## A. One Original image with n-1 Randomly Chosen Gray images (M1)

In this method, one original Biometric image $S_1$ and *n*-1 different Gray images $C_1, C_2,...,C_3$ are used for generation of Cancelable Biometric template. For experiment works, one original Biometric image and three natural gray scale Cover images are chosen. The process of Secret Shares generation comprised of following steps: (i) Generation of Temporary Shares $TS_1, TS_2,...,TS_n$ by XORing Secret Biometric $S_1$ image with Cover images $C_1, C_2,...,C_{n-1}$ (ii) Noisy Shares $NS_1, NS_2,...,NS_n$ are generated from Temporary Shares. (iii) Left Bit-wise Reversal of individual bit of Noisy Shares. The complete algorithm of above steps is explained in TABLE II. A well-structured diagram of original Secret Biometric image, Temporary Shares, intermediate Noisy Shares and Secret Shares is depicted in Fig. 2. The only advantage of this method is its storage scheme, which is based on Visual Secret Share technique. The main drawback of this method is that the Secret Shares reveal some coarse structure of the original Biometric as shown in Fig. 2.

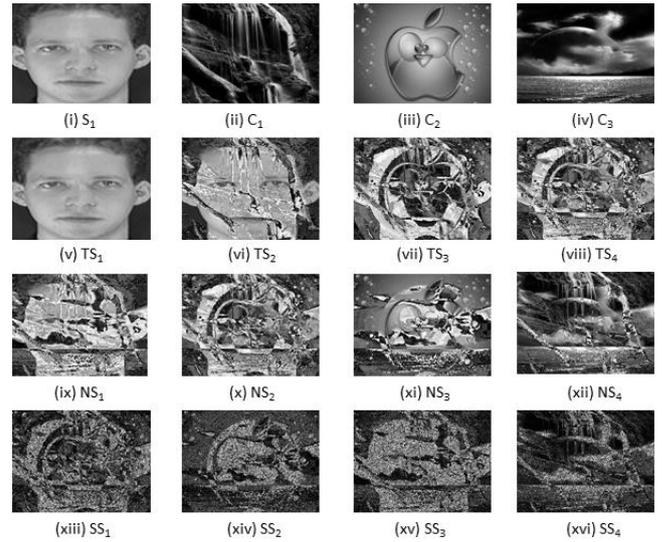

Fig. 2. step-by-step process of *M1* method: (i) Secret image, (ii) — (iv) Cover images, (v) — (viii) Temporary Shares, (ix) — (xii) Noisy Shares, (xiii) — (xvi) Secret Shares.

## B. One Original image with n-1 Random Permuted version of Original image (M2)

In first method, Cancelable Biometric Secret Shares are generated using one original Biometric image $S_1$ and *n*-1 Gray Cover images $C_1, C_2$ and $C_3$. In this method, Cover images are randomly permuted version of original image. The main objective of any Cancelable Biometric based technique is that, its Cancelable Biometric templates should not leak any partial or full information regarding its original Biometric image. The major difference between *M2* with *M1* is, Cover images are randomly permuted version of the original Secret image i.e. Cover images $C_1, C_2,...,C_3$ are randomly permuted version of $S_1$. While in first method, Cover images are randomly chosen gray images. Rest all the Shares generation process of Temporary Shares, Noisy Shares and Secret Shares is similar with method *M1*. For Experiments, one original Biometric image and three Cover images are chosen, and these Cover images are randomly permuted version of original Biometric image. The major drawback of this method is its Secret Share $SS_1$ reveals partial information about original Biometric information as can be seen from Fig. 3.

## C. All n images are Random Permuted version of Original image (M3)

The main similarity between this method and above two methods is that, it is also using one Secret image $S_1$ and $n$-1 Cover images $C_1, C_2,…,C_{n-1}$. The major difference of this method with above two methods is that, both Secret image $S_1$ and $n$-1 Cover images are randomly permuted version of original Biometric image S as shown in Fig. 4. It means whole Algorithm is applied on randomly permuted version of original Secret image S. For experimental work, one Secret image and three Cover images are chosen, Both Secret image and Cover images are randomly permuted version of the original Biometric image. Rest all the functioning is similar with method *M1* and *M2* i.e. Temporary Shares, Noisy Shares & Secret Shares generation steps. As we can see from Fig. 4, no Secret Share reveals any information about original Biometric.

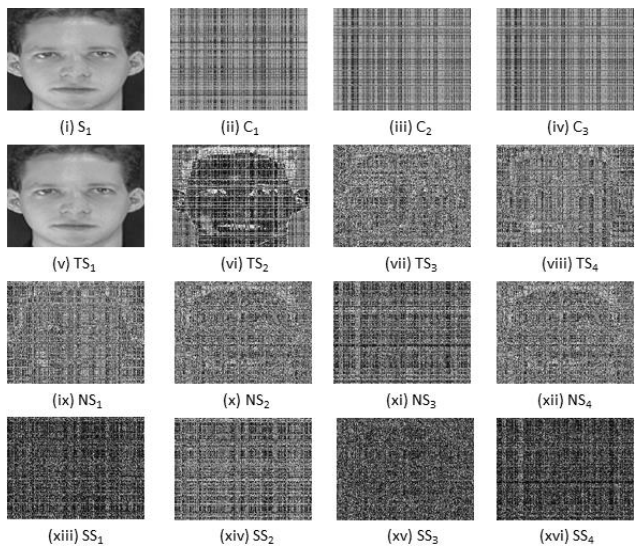

Fig. 3. step-by-step process of *M2* method: (i) Secret image, (ii) — (iv) Cover images, (v) — (viii) Temporary Shares, (ix) — (xii) Noisy Shares, (xiii)—(xiv) Secret Shares.

Major advantages of this technique over other two methods are (i) its Secret Shares do not reveal any partial and full information about the original Biometric image (ii) If anyhow an intruder accessed all Secret Shares, after applying reverse engineering process if Secret image $S_1$ and $n$-1 Cover images are retrieved, which are again randomly permuted version of original Secret image (distorted). Hence it proofs the non-invertible property of Cancelable Biometric based application.

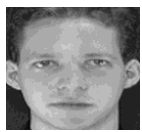

Fig. 4. Original Biometric image (Secret)

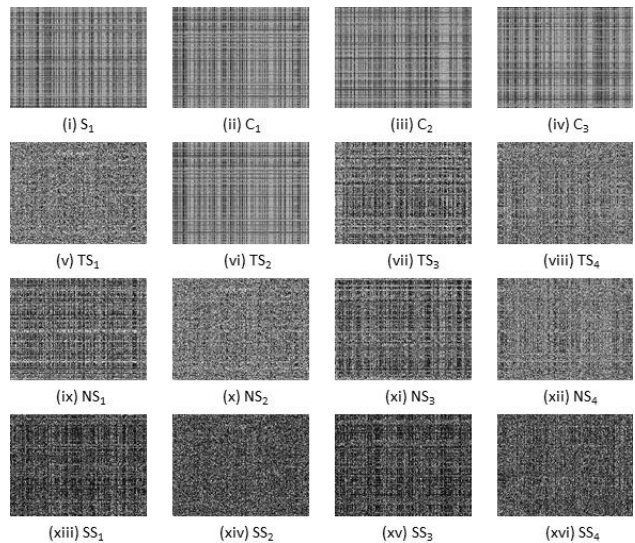

Fig. 5. Step-by-step process of *M3* method: (i) Secret image, (ii)—(iv) Cover images, (v)—(viii) Temporary Shares, (ix)—(xii) Noisy Shares, (xiiii)—(xvi) Secret Shares.

## IV. RESULTS AND DISCUSSION

Experimental works have performed on one Secret image an three Cover images. Publically available ORL Face database and IIT Delhi Iris database are used for three proposed methods. ORL Face image has dimension 112×94 with .pgm image format, while IIT Delhi Iris database image has dimension 320×240 with .bmp image format. The result of all three proposed methods is shown in TABLE V. Our result's TABLE consists with different performance measures. If two images are same, each measure has an ideal value as shown in TABLE V. *Correlation Coefficient* value 1 indicates that two images are same. It can be observed from Table V that the average correlation between two image approaches 0 which shows that the original Biometric image and the Cancelable template generated using any of the three methods are not related and hence can be treated as independent. *RMSE* value 0 indicates that two images are same. However, the value computed using the proposed methods is different from 0. *PSNR* value ranges between 0 to ∞ for two images being different or same respectively. The proposed methods have a value near to 0 than ∞. *SSIM* value range between -1 and +1 different and same images respectively. In our experiments, we have obtained *SSIM* value approximately 0, which shows total distortion in images. For two same images *MAE, NPCR, UNCI* values 0 indicates that two images are same. For our proposed methods, these performance measures achieve values other than 0 which indicated that the original Biometric image and the Cancelable Biometric templates are different. Overall, the performance of all the proposed methods is good; however, the method *M3* achieves best results qualitatively and quantitatively on ORL dataset except SSIM. On Iris dataset, methods M2 and M3 are close competitors and M2 is better in terms of RMSE, MAE and PSNR while M3 is better in rest.

TABLE IV.

**Formulas for Performance measures**

| | *Formula* | *Description* |
|---|---|---|
| *Correlation Coefficient (Cr)* | $C_r = \dfrac{\sum_m \sum_n (I_{mn} - \bar{I})(S_{mn} - \bar{S})}{\sqrt{\left(\left(\sum_m \sum_n (I_{mn} - \bar{I})\right)^2\right) - \left(\left(\sum_m \sum_n (S_{mn} - \bar{S})\right)\right)^2}}$ | *I* and *S* are images |
| *MSE (Mean Square Error)* | $MSE = \dfrac{1}{W \times H} \sum_{i=1}^{W} \sum_{j=1}^{H} (I(i,j) - S(i,j))^2$ | $I(i,j) - S(i,j)$ is error and $W \times H$ is size of image |
| *Root Mean Square Error (RMSE)* | $\sqrt{MSE}$ | MSE is Mean Square Root |
| *Peak signal to noise ratio (PSNR)* | $PSNR(dB) = 20 \log_{10} \dfrac{255}{\sqrt{MSE}}$ | 255 is the maximum gray value |
| *Structural Similarity (SSIM)* | $SSIM(I,S) = \dfrac{(2\mu_I \mu_S + C_1)(2\sigma_{IS} + C_2)}{(\mu_I^2 + \mu_S^2 + C_1)(\sigma_I^2 + \sigma_S^2 + C_2)}$ | $\mu_I, \mu_S, \sigma_I, \sigma_S, \sigma_{IS}$ are mean and variance and covariance of image *I* and *S*, $C_1$ and $C_2$ are constants |
| *Mean Absolute Error (MAE)* | $MAE = \dfrac{1}{W \times H} \sum_{i=1}^{W} \sum_{j=1}^{H} |I(i,j) - S(i,j)|$ | $I(i,j) - S(i,j)$ is error |
| *No. of Pixel Change Rate (NPCR)* | $NPCR(I_i, S_j) = \dfrac{\sum_{i,j} D(i,j)}{W \times H} \times 100\%$ | (i) $D(i,j) = 0$; if $I(i,j) = S(i,j)$<br>(ii) $D(i,j) = 1$; if $I(i,j) \neq S(i,j)$ |
| *Unified Average Change Intensity (UACI)* | $UACI = \dfrac{1}{W \times H} \times \dfrac{\sum_{i,j} |I(i,j) - S(i,j)|}{255} \times 100\%$ | 255 is maximum value of gray image. |

TABLE V.

**Performance of the Proposed Methods on ORL (Face) and IIT Delhi (Iris) Datasets**

| Database | Ideal Values | Ideal Value to Show Two Images are Same | | | | | | |
|---|---|---|---|---|---|---|---|---|
| | | *Cr* | *MSE* | *MAE* | *PSNR* | *SSIM* | *NPCR* | *UACI* |
| | | **1.00** | **0** | **0** | **∞** | **1.00** | **0** | **0** |
| | | *Cr* | *MSE* | *MAE* | *PSNR* | *SSIM* | *NPCR* | *UACI* |
| ORL | Method 1 | 0.1353 | 7971.40 | 44.27 | 9.220 | 0.0389 | 99.09 | 17.57 |
| | Method 2 | -0.0207 | 11026.35 | 59.24 | 7.724 | -0.0071 | 99.28 | 23.49 |
| | Method 3 | -0.0276 | 11110.34 | 59.54 | 7.691 | -0.0157 | 99.25 | 23.66 |
| IIT Delhi | Method 1 | 0.0719 | 10033.71 | 51.54 | 8.368 | 0.0339 | 98.83 | 20.78 |
| | Method 2 | -0.0155 | 12710.92 | 68.71 | 7.154 | -0.0048 | 64.11 | 27.18 |
| | Method 3 | -0.0192 | 11026.75 | 59.15 | 7.721 | -0.0039 | 99.22 | 23.39 |

## V. CONCLUSION AND DISCUSSION

We have proposed three different methods *M1, M2 and M3* for generating Cancelable Biometric templates. For measuring the performance of these methods, we have used different measures as listed in Table IV. From our experimental work, we have concluded that *M3* is more secure among all the three methods as shown by Figs. 2, 3, 4 and resultant TABLE V. In our future work, we shall explore more sophisticated methods which are more secure and non-invertible in nature.

ACKNOWLEDGMENT

We acknowledge Ministry of Human Resource Development, Govt. of India for supporting this research by providing fellowship to one of the authors, Ms. Manisha. One of the authors, Dr. Nitin Kumar is thankful to Uttarakhand State Council for Science and Technology, Dehradun, Uttarakhand, India for providing financial support towards this research work (Sanction No. UCS & T/R & D-05/18-19/15202/1 dated 28-09-2018)

## REFERENCES

[1] Manisha and N. Kumar. , " Cancelable Biometrics: A Comprehensive Survey," Artif Intell Rev. Springer. Netherlands, 2019. https://link.springer.com/article/10.1007/s10462-019-09767-8

[2] M. Deshmukh, N. Nain, and M. Ahmed, " Efficient and secure multi secret sharing schemes based on boolean XOR and arithmetic modulo," Multimedia Tools and Applications, vol. 77(1), pp. 89-107, 2016.

[3] N. Kumar and M. Rawat, " RP-LPP: a random permutation based locality preserving projection for Cancelable Biometric recognition," Multimedia Tools and Applications, pp. 1-19, 2019. https://doi.org/10.1007/s11042-019-08228-2

[4] L. Bai, "A reliable (k, n) image secret sharing scheme," in IEEE International Symposium on Dependable, Autonomic and Secure Computing, vol.(2), pp. 31-36, September 2006.

[5] S. Zou, Y. Liang, , L. Lai and S. Shamai, "An information theoretic approach to secret sharing," IEEE Transactions on Information Theory, vol. 61(6), pp. 3121-3136, 2015.

[6] N. Kumar, S. Singh and A. Kumar, "Random permutation principal component analysis for cancelable biometric recognition," Applied Intelligence 48, pp. 2824-2836, 2018. https://doi.org/10.1007/s10489-017-1117-7

[7] H. Kaur and P. Khanna, " Random Distance Method for Generating Unimodal and Multimodal Cancelable Biometric Features," IEEE Transactions on Information Forensics and Security, vol 14(3), pp. 709-719, 2019.

[8] N. Lalithamani and K. P. Soman, " Towards generating irrevocable key for cryptography from cancelable fingerprints," In IEEE International Conference on Computer Science and Information Technology, pp. 563-568, August 2009.